\documentclass[runningheads]{llncs}

\usepackage{eccv}

\usepackage{eccvabbrv}

\usepackage{graphicx}
\usepackage{booktabs}

\usepackage[accsupp]{axessibility}  

\usepackage{hyperref}

\usepackage{orcidlink}

\begin{document}

\title{LSD3K: A Benchmark for Smoke Removal from Laparoscopic Surgery Images} 

\titlerunning{Abbreviated paper title}

\author{Wenhui Chang,
Hongming Chen }

\authorrunning{F.~Author et al.}

\institute{College of Electronic Information Engineering, Shenyang Aerospace University }

\maketitle

\begin{abstract}
  Smoke generated by surgical instruments during laparoscopic surgery can obscure the visual field, impairing surgeons' ability to perform operations accurately and safely. Thus, smoke removal task for laparoscopic images is highly desirable. Despite laparoscopic image desmoking has attracted the attention of researchers in recent years and several algorithms have emerged, the lack of publicly available high-quality benchmark datasets is the main bottleneck to hamper the development progress of this task. To advance this field, we construct a new high-quality dataset for Laparoscopic Surgery image Desmoking, named LSD3K, consisting of 3,000 paired synthetic non-homogeneous smoke images. In this paper, we provide a dataset generation pipeline, which includes modeling smoke shape using Blender, collecting ground-truth images from the Cholec80 dataset, random sampling of smoke masks and etc. Based on the proposed benchmark, we further conducted a comprehensive evaluation of the existing representative desmoking algorithms. The proposed dataset is publicly available at \url{https://drive.google.com/file/d/1v0U5_3S4nJpaUiP898Q0pc-MfEAtnbOq/view?usp=sharing}.
\end{abstract}

\section{Introduction}
\label{sec1}
Image desmoking is a crucial research topic in minimally invasive surgery, aiming to enhance visual clarity during laparoscopic procedures by mitigating the obscuring effects of smoke generated from electrosurgical devices \cite{hong2023mars,su2023multi}. Recently, this field has received increasing attention as it addresses a persistent challenge in laparoscopy, improving surgical precision and minimizing potential risks associated with impaired visibility. Thus, smoke removal task for laparoscopic images is highly desirable.

In fact, laparoscopic image desmoking is more challenging than natural image dehazing. That is because the dynamic and unpredictable nature of laparoscopic smoke underscores its inherent non-uniformity and randomness compared to outdoor fog \cite{zhang2023progressive}. When we revisit the development in this field, researchers have primarily directed their focus on the algorithm design, while relative fewer attention has been paid on the benchmark dataset. The lack of publicly available high-quality benchmark datasets is the main bottleneck to hamper the development progress of this task. Thus, it is urgent to construct a benchmark dataset and a baseline for addressing non-homogeneous smoke removal for laparoscopic surgery scenes. 
\begin{figure*}[!t]
\centering
\includegraphics[width=1.0\textwidth]{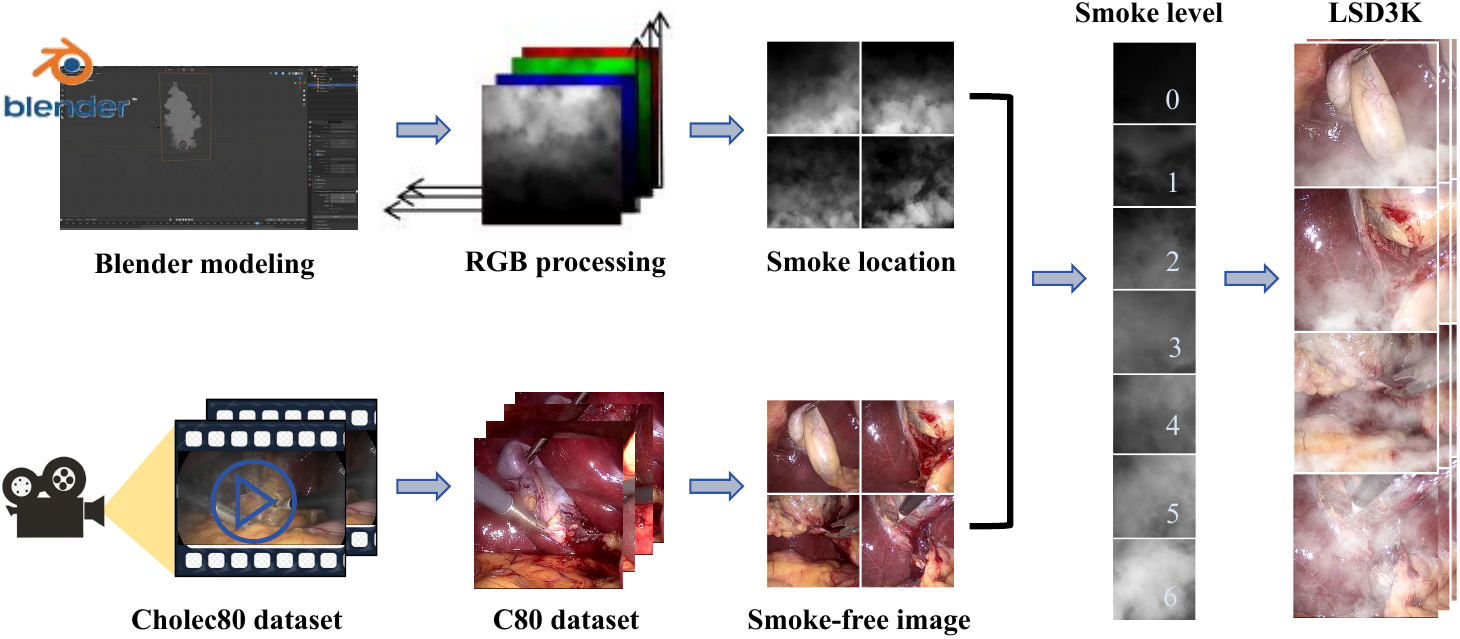}
\caption{Illustration of the dataset generation pipeline.}
\label{fig1}
\end{figure*}
		
For data-driven methods, obtaining paired smoke and smoke-free images is not easily feasible. Constructing simulated paired matching smoke datasets is a costly and time-consuming task. Researchers often extract smoke-free images from publicly available surgical videos and then linearly overlay synthesized smoke as preprocessing for network training. In previous works concerning the synthesis of datasets for endoscopic procedures, there has not been a fully unified standard for selecting background images and synthesizing smoke. Especially with medical datasets, they not only consume valuable medical resources but also require high levels of accuracy and quantity to meet medical practice standards. Obtaining images with thousands of different density smoke masks is not an easy task. Given the requirement to meet medical practice standards, manually acquiring density masks and annotations for numerous image pairs in a real-world dataset seems impractical \cite{chen2023towards,chen2024towards,chang2024uav}. Synthetic datasets offer a straightforward and scalable alternative to manually annotating images. Given the current practical challenges, there is a significantly increased demand for synthetic smoke datasets to address these issues.

To this end, we construct a new dataset called LSD3K, and provide a dataset generation pipeline for laparoscopic image desmoking, consisting of 3,000 paired synthetic smoky images. Furthermore, based on this dataset, we conducted a comprehensive evaluation of several advanced image desmoking algorithms. These methods were assessed quantitatively and qualitatively using our new dataset. Our evaluation and analysis highlighted the performance and limitations of existing methods and stimulated further research into more robust algorithms. The proposed LSD3K dataset is publicly available for research purposes. We believe that this work can provide new insights into medical image data synthesis. 

The rest of this paper is structured as follows. We review the related work in the field of smoke removal from laparoscopic surgery images in Section~\ref{sec2}. In Section~\ref{sec3}, we provide a detailed description of the pipeline used to construct the dataset.In Section~\ref{sec4}, we analyze the performance of existing algorithms in the benchmark test. The discussion is presented in Section~\ref{sec5}. Finally, the concluding remarks will be given in Section \ref{sec6}.

\section{Related Work}
\label{sec2}
To our knowledge, there have been few recent works on image-based smoke removal in laparoscopic settings \cite{wang2018variational,chen2019smokegcn,wang2019multiscale,salazar2020desmoking,venkatesh2020unsupervised,zhou2022synchronizing,su2023multi,zhang2023progressive}. Simultaneously, due to the specificity of minimally invasive surgical procedures, acquiring paired real surgical smoke datasets for deep learning is nearly impossible. The development of smoke removal algorithms is hindered by the difficulty in constructing large-scale simulated paired matching datasets. In this section, we focus on the generation of datasets specifically tailored for smoke removal methods applied to endoscopic images.

Wang et al. \cite{wang2018variational} proposed an efficient variational-based smoke removal method for laparoscopic images. The performance of the proposed method was quantitatively and qualitatively evaluated using two publicly available real smoked laparoscopic datasets and one generated synthetic dataset. The real smoked laparoscopic datasets were obtained from the Hamlyn Centre laparoscopic/endoscopic video dataset page \cite{ye2017self}. The synthetic dataset was generated by utilizing Berlin noise \cite{bolkar2018deep} to produce synthetic smoke, which was then linearly embedded into artificially selected ground truth smoke-free images. In \cite{chen2019smokegcn}, the authors have developed a novel generative collaborative learning approach called DesmokeGCN. The algorithm utilizes real laparoscopic images obtained from the Hamlyn Centre laparoscopic video dataset \cite{ye2017self} and the Cholec80 dataset \cite{zhang2023progressive} as background images. Additionally, it employs the 3D rendering engine Blender for synthesizing non-uniform smoke. In \cite{wang2019multiscale}, Wang et al. further proposed a real-time smoke removal method based on Convolutional Neural Networks (CNNs). They manually selected 100 smoke-free images from the Hamlyn Centre laparoscopic video dataset \cite{ye2017self} and used a dataset consisting of synthetic smoke images generated by Blender and Adobe Photoshop to train the network. In \cite{zhou2022synchronizing}, Zhou et al. proposed a new method named Dessmoke-CycleGAN. Smoke and smoke-free images used in the experiments were captured from da Vinci surgical robot videos. Additionally, random smoke generated by Blender was linearly added to smoke-free images for training and testing purposes. Based on the above, there is an urgent need to construct a high-quality paired dataset to address the non-uniform smoke removal issue in laparoscopic surgical scenes.

\begin{figure*}[!t]
\centering
\includegraphics[width=1.0\textwidth]{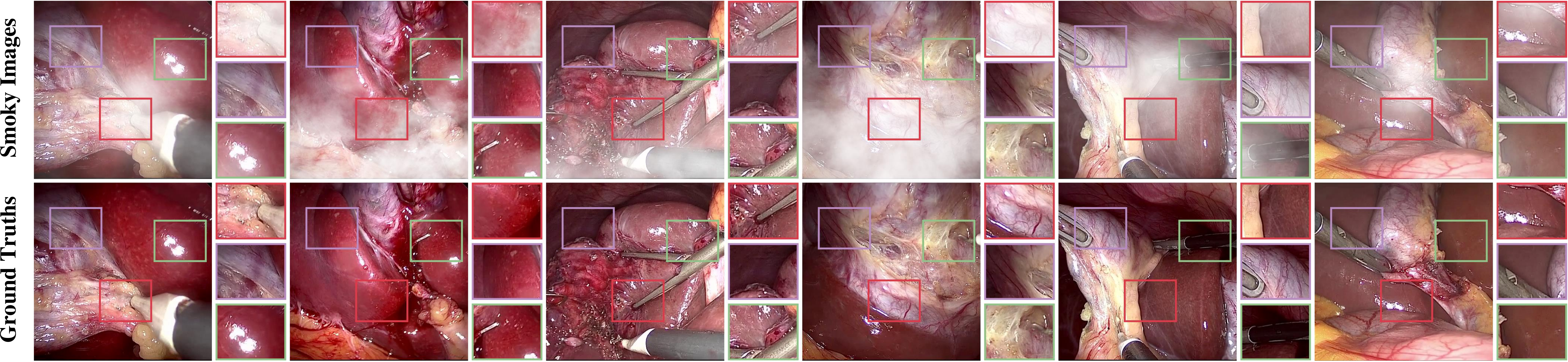}
\caption{More example image pairs sampled from the proposed LSD3K. Best viewed by zooming in the figures.}
\label{fig2}
\end{figure*}

\section{Dataset Construction}
\label{sec3}
While there are several large-scale real endoscopic surgery datasets available, they are limited by the constraints of actual surgical environments and lack diversity in smoke, rendering them unsuitable for training and testing deep learning networks. In this section, we provide a detailed overview of the synthesis process in LSD3K.
\begin{figure}[!t]
\centering
\includegraphics[width=1.0\columnwidth]{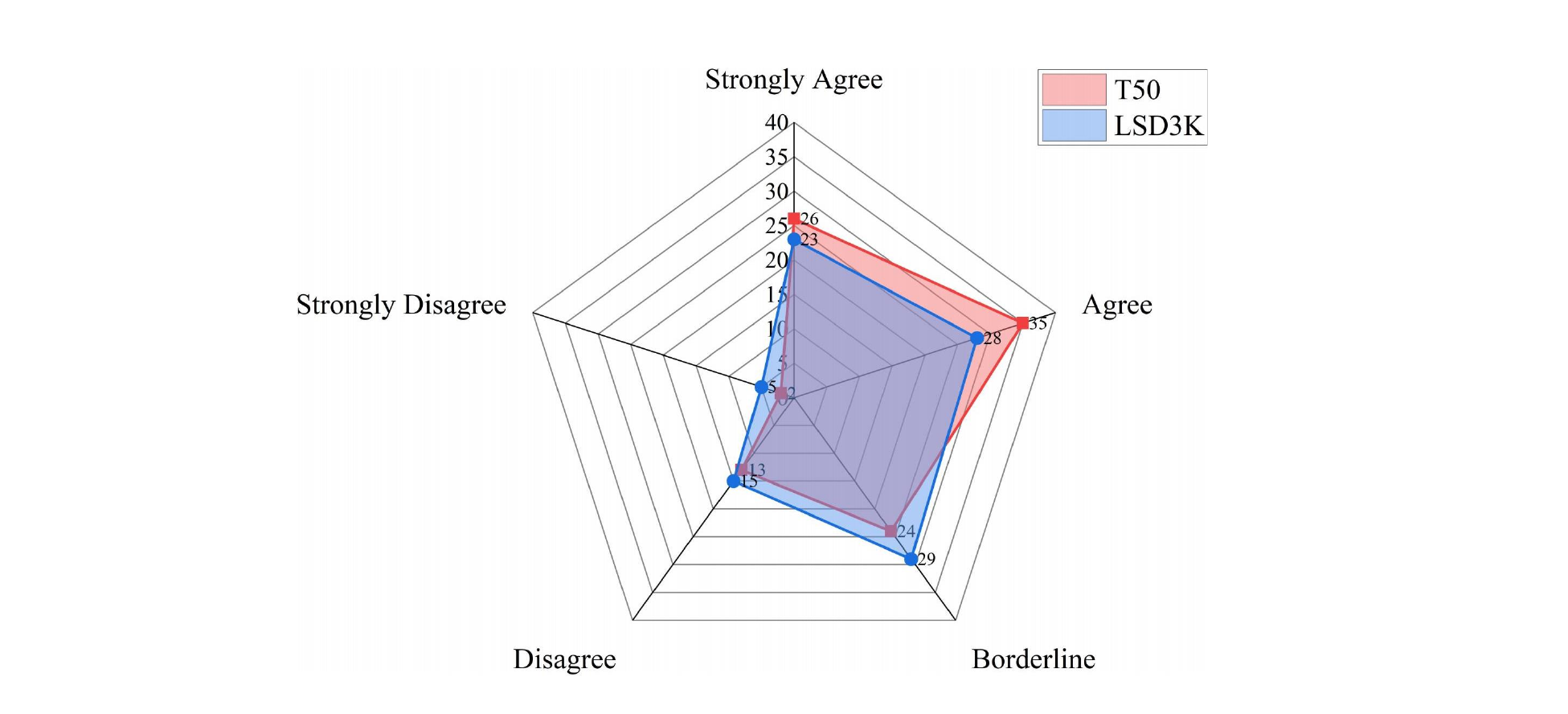}
\caption{User study results. The ratings given by all participants on different smoke datasets.}
\label{fig3}
\end{figure}

\begin{figure*}[!t]
\centering
\includegraphics[width=1.0\textwidth]{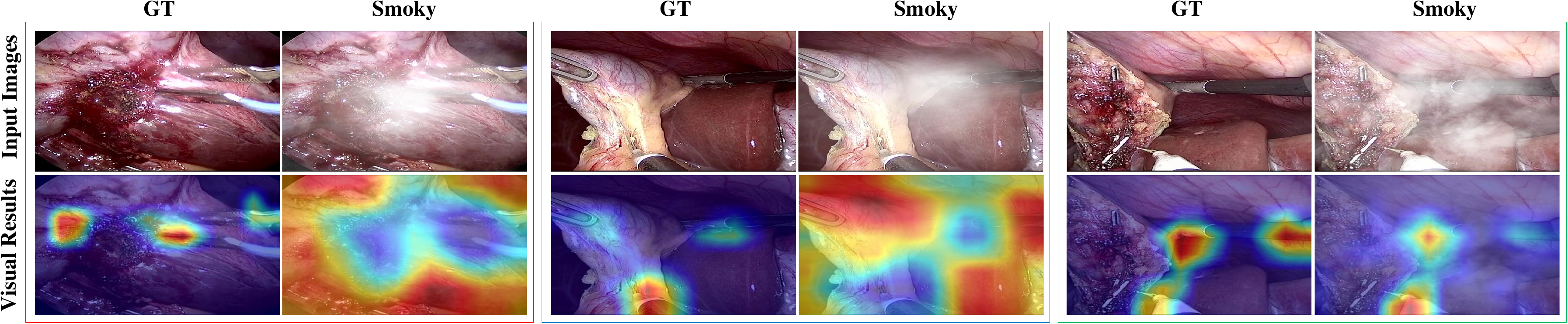}
\caption{Visual results of paired image instrument tracking.}
\label{fig4}
\end{figure*}

\subsection{Smoke Synthesis}
\label{sec:3.1}
The realistic simulation of heterogeneous smoke is crucial for training and testing models developed. Due to the typically narrow field of view of endoscopes during surgical procedures, and the fact that smoke generated from procedures such as electrocautery and laser ablation is random and localized, it is unrelated to depth. Traditional haze models \cite{cai2016dehazenet,tang2014investigating} and Berlin noise functions \cite{bolkar2018deep} are not designed for image desmoking and cannot address the specific characteristics of smoke \cite{pei2019single}. Furthermore, the dataset images generated by these methods overly simplify the distribution of smoke, lacking the ability to express complex scenes. Particularly, they do not consider the non-uniformity of smoke, which is a common distribution pattern in smoke during endoscopic surgery. Furthermore, in laparoscopic images, the light source is provided by unevenly distributed instruments, and the organ surfaces are not Lambertian surfaces\cite{cai2016dehazenet}. To address this issue, we employ Blender, an open-source 3D creation software, to synthesize simulated endoscopic surgery smoke images for training. Modern rendering engines in Blender use sophisticated, physics-based built-in models, providing realistic and diverse smoke shapes and densities \cite{huang2021single}. This effectively addresses the non-uniformity of smoke, and the advantages of using such an approach are evident. Here, we provide a detailed description of the synthesis process. Smoke ${I}_{smoke}$ can be defined as:
\begin{equation}
{I}_{smoke}(x,y)=Blender({I}_{rand},{D}_{rand},{P}_{rand})
\end{equation}
where ${I}_{rand}$ denotes smoke intensity, ${D}_{rand}$ represents smoke density, ${P}_{rand}$ signifies the starting position of smoke, The Intensity ${I}_{rand}$ stands for the smoky solid particles transferred at a certain degree, The Density ${D}_{rand}$ indicates the non-uniform diffusion of smoke-like solid particles within a specific volume, and the Position ${P}_{rand}$ represents the general starting position of smoke within the image area. 

As graphics are typically of the color type, the smoke mask ${I}_{mask}$ is derived from the brightness of the R, G, B channels in the original smoke ${I}_{smoke}$, which can be defined as: 
 \begin{equation}
 	\begin{aligned}
{I}_{mask}(x,y)=(0.3*{I}_{smoke}{(x,y)}^{R})\\+(0.59*{I}_{smoke}{(x,y)}^{G}+(0.11{I}_{smoke}{(x,y)}^{B}).
 		\end{aligned}
 	\end{equation}

By overlaying smoke images with the same density, intensity, and position on a smoke-free image, a smoke image can be obtained:
\begin{equation}
{I}_{smoked-image}(x,y)={I}_{smoke-free}(x,y)+{I}_{mask}.
\end{equation}
The randomness in the rendering process helps avoid overfitting of the network and allows for the generation of a sufficient number of synthetic smoke images for training. These images incorporate smoke masks with various positions and smoke levels added using a 3D graphics engine. With the aid of a powerful rendering engine, we are capable of synthesizing an unlimited number of realistic images simulating surgical smoke for network training. The smoke density is graded, ranging from 0 to 6, with 0 defined as smoke-free and 6 representing the maximum smoke density in the generated smoke images. Figure~\ref{fig1} provides a detailed illustration of the dataset generation process and the distribution of smoke density levels.

\subsection{Dataset Statistics}
\label{sec:3.2}
To prepare the synthetic data, we obtain clear background images from the publicly available dataset C80 \cite{twinanda2016endonet}, which comprises images from the Cholec80 dataset \cite{zhang2023progressive}. Cholec80 consists of 80 cholecystectomy videos performed by 13 surgeons. Among these, we utilize the variance of the Laplacian function \cite{ye2017self} for image selection, followed by a second round of manual inspection to ensure the absence of surgical smoke in the images, ensuring the ground truth. Finally, we collect 660 clear and smoke-free source images. Subsequently, we linearly add six different densities (opacity levels) of synthesized random smoke. After confirmation, smoke of various densities, intensities, and positions are added, resulting in the generation of a diversified endoscopic surgery smoke dataset. In the end, we have generated 3000 pairs of images. We randomly select 200 pairs of synthesized smoke images for testing the network and 2800 pairs for training. Additionally, for ease of validating the effectiveness of training the network for smoke removal, we includ 50 real endoscopic surgery smoke images in the test set. These real images are sourced from the publicly available dataset CholecT50 \cite{nwoye2022rendezvous}. This dataset is referred to as LSD3K. Furthermore, for experimental convenience, the resolution of all synthesized images in LSD3K is uniformly cropped to $480 \times 480$ pixels. In Figure~\ref{fig2}, we provide a detailed presentation of the dataset synthesis process.
\begin{table*}[t]
\centering
\caption{Quantitative comparisons on the LSD3K benchmark dataset. ``\#FLOPs'' and ``\#Params'' represent FLOPs (in G) and the number of trainable parameters (in M), respectively.}
\resizebox{0.95\textwidth}{!}{
\begin{tabular}{cccccccc}
\hline
Methods   & Input  & DCP\cite{he2010single}   & AOD-Net\cite{li2017aod} & GridDehazeNet\cite{liu2019griddehazenet} & Restormer\cite{zamir2022restormer} &
Dehamer\cite{guo2022image} & DehazeFormer\cite{song2023vision} 
\\ \hline
Category  & -      & Prior  & CNN    & CNN  & Transformer  & Transformer & Transformer  \\
PSNR      & 15.122  & 18.46  & 16.81  &32.45   &25.10  &26.73   &29.72          \\
SSIM      & 0.7795 & 0.8637 & 0.8249 &0.9780  & 0.9344  &0.9263 & 0.9665       \\
LPIPS   & 0.2247   & 0.1552    & 0.2085  &0.0290  &0.0755 &0.0843  &0.0403      \\
\#Params  & -      & -      & 0.02   & 0.96   & 26.13       & 132.45       & 4.64       \\ \hline
\end{tabular}
}
\label{table1}	
\end{table*}
\begin{figure*}[!t]
\centering
\includegraphics[width=1.0\textwidth]{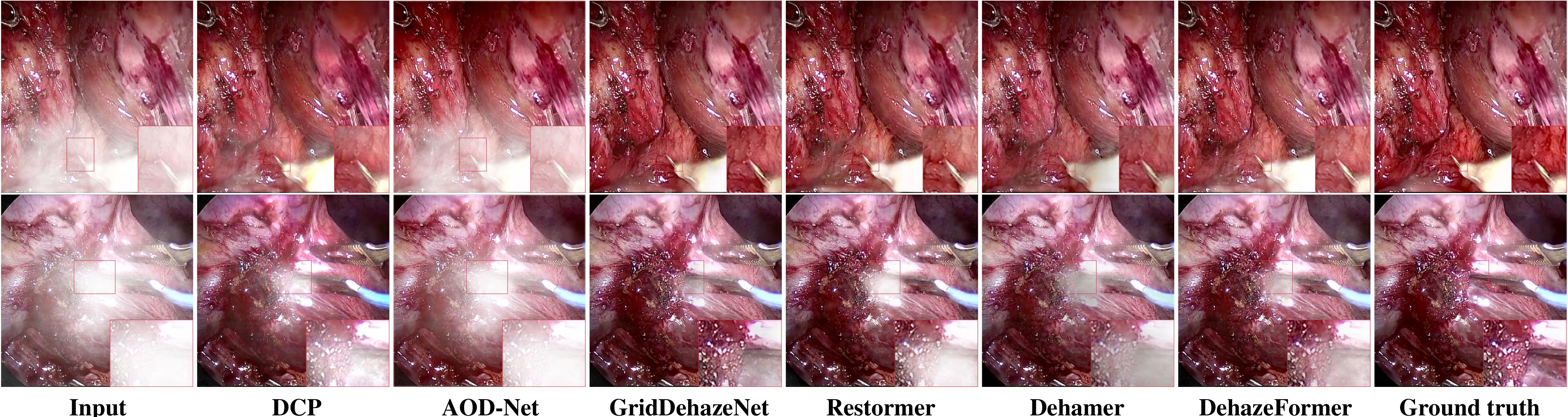}
\caption{Visual results of paired image instrument tracking.}
\label{fig5}
\end{figure*}

\subsection{User Study}
\label{sec:3.3}
Next, we conduct a human subjective survey to assess the quality of the synthesized smoke images. Here, we randomly select 20 synthesized smoke images from the LSD3K dataset as one group of samples, while the other group of 20 smoke images are sourced from the real endoscopic surgery dataset (CholecT50 \cite{nwoye2022rendezvous}). We recruit 20 participants, including 10 volunteers from the medical surgery industry and 10 volunteers from non-medical sectors. Each participant is shown 10 smoke images randomly selected from both samples (samples one and two had an equal number of images). Then, employing a 5-point Likert scale (\ie, strongly agree, agree, borderline, disagree, strongly disagree), all participants is asked to assess the perceived realism of each image. Finally, we receive 200 ratings for the two sets of samples. Despite the relatively small number of evaluators, we observe a strong consensus and minimal inter-rater differences in the ratings for the same paired comparison results. This indicates that the ratings are reliable.

To make the differences in ratings between the two sets of samples more intuitive, the radar chart of rating levels is depicted in Figure~\ref{fig3}. Although LSD3K still falls short in terms of ratings with strong agreement compared to real smoke images, in terms of overall ratings, the rating distributions of the two groups nearly overlap. This may indicate that LSD3K's visual realism is approaching that of real smoke images.
\subsection{Application}
\label{sec:3.4}
Surgical instrument segmentation is a crucial task that can significantly impact the outcomes of medical procedures \cite{theckedath2020detecting}. To investigate the necessity of smoke removal processes for downstream vision-based medical surgical applications, we further evaluated unprocessed smoke images. For surgical instrument tracking, we applied a popular instrument tracking network model (ResNet50 \cite{chmarra2007systems}) to assess the influence of smoke on visual images during the surgical process. Figure~\ref{fig4} illustrates the visualization of instrument detection results for three pairs of synthetic smoke images in LSD3K. It can be observed that all smoke-containing images exhibit varying degrees of interference with detection accuracy, particularly evident during instances of dense smoke during surgical procedures. Detection accuracy of surgical instruments on smoke-free background images is relatively high, with good attention to the instruments in the visualized images. However, the generation of smoke affects semantic information, resulting in more areas of misclassification. Thus, combining the visualization results, it is evident that smoke generated during endoscopic surgery can impact normal medical procedures, highlighting the necessity for a synthetic endoscopic smoke dataset.

\section{Algorithm Benchmarking }
\label{sec4}
In this section, based on the newly proposed benchmarks, we evaluated four representative algorithms: DCP\cite{he2010single}, AOD-Net\cite{li2017aod}, GridDehazeNet\cite{liu2019griddehazenet}, Restormer\cite{zamir2022restormer}, Dehamer\cite{guo2022image} and DehazeFormer\cite{song2023vision}. To ensure a fair comparison, we utilize the official released codes of these methods. Each method underwent retraining for the LSD3K benchmark tests on servers equipped with NVIDIA RTX 4090 GPUs.
\subsection{Quantitative Evaluation}
\label{sec:4.1}
Table~\ref{table1} presents the quantitative performance evaluation results of various algorithms on the LSD3K dataset. To assess the quality of the demosaiced images, three quantitative evaluation metrics were used: PSNR, SSIM, and LPIPS\cite{zhang2018unreasonable}. From the results in Table~\ref{table1}, it is evident that GridDehazeNet\cite{liu2019griddehazenet} achieved the best quantitative results in smoke removal performance. However, its model complexity is relatively higher compared to traditional CNN methods. Among the transformer-based desmoking algorithms, DehazeFormer\cite{song2023vision} achieved the highest scores. To comprehensively evaluate the performance and efficiency of different algorithms, future research can delve deeper into strategies that maintain high performance while reducing model complexity, to meet the resource constraints of medical equipment.
\subsection{Qualitative Evaluation}
\label{sec:4.2}
Figure~\ref{fig5} illustrates the visual comparison results of different baseline algorithms on our proposed benchmark. It is evident from the figure that the traditional image processing algorithm DCP\cite{he2010single} has limitations in its effectiveness and leads to visual distortions.The AOD-Net\cite{li2017aod}, which is based on the atmospheric scattering model and treats haze as a uniform medium, also shows suboptimal desmoking results. In contrast, both GridDehazeNet\cite{liu2019griddehazenet}, 
and DehazeFormer\cite{song2023vision}  demonstrate the best visual performance, effectively removing haze while minimizing pixel distortion.
\section{Discussion}
\label{sec5}
For decades, with the success of deep learning algorithms, the research community in image processing and computer vision has been addressing general image dehazing and smoke removal tasks, ranging from acquiring clear outdoor scenes affected by weather conditions to restoring surgical scenes. However, several challenges persist.In this task, collecting paired data is difficult, if not impractical. In Section~\ref{sec2}, we summarized that past researchers typically extracted smoke-free images from publicly available surgical videos and then linearly superimposed synthesized smoke as a preprocessing step for network training. However, these datasets suffer from domain gaps between synthetic smoke and real-world smoke, especially in some dense smoke images. To address current real-world challenges, there is a significantly increased demand for synthetically high-quality endoscopic smoke datasets.

The novelty of this work lies in bringing smoke removal in surgical images into the realm of real-world applications, which holds greater practical significance. Training networks with synthesized smoke datasets addresses the deficiency in training data for medical applications and bridges the significant gap between simulation and reality.For instance, LSD3K draws backgrounds from various laparoscopic and endoscopic videos, exhibiting diverse image colors and tones. Smoke is rendered by a 3D rendering engine using random intensities, densities, textures, and positions. This addresses the challenging issue of deep learning's reliance on labor-intensive manual annotation of ground truth training data, particularly for medical datasets where domain expertise is crucial in annotation. Additionally, LSD3K holds many potential applications in surgical human-machine interaction.

\section{Conclusion}
\label{sec6}
In this paper, we have proposed a new high-quality dataset for smoke removal from laparoscopic surgery images. We provid a detailed overview of the synthesis process for the LSD3K dataset, including modeling smoke shape using Blender, collecting ground-truth images from the Cholec80 dataset, random sampling of smoke masks and etc. Based on the proposed dataset, we provide new insights into medical image data synthesis and call on researchers to further focus on this field and propose more robust algorithms.

%
%
\bibliographystyle{splncs04}
\bibliography{main}
\end{document}